\documentclass[twocolumn,showpacs,preprintnumbers,amsmath,amssymb]{revtex4}
\input xy
\xyoption{all}
\usepackage{amsfonts,latexsym}
\begin{document}
%\renewcommand{\thefootnote}{\fnsymbol{footnote}}
%\draft
\title{
Quark confinement without a confining force
} 

\author{P.S. Isaac}  
\email{isaac@ms.u-tokyo.ac.jp} 
\affiliation{Graduate School of Mathematical Sciences, The
University of Tokyo, 3-8-1 Komaba, Meguro, 153-8914,
Tokyo, Japan} 
\author{W.P. Joyce} 
\email{william.joyce@canterbury.ac.nz} 
\affiliation{Department of Physics and Astronomy, The University of
Canterbury, Private Bag 4800, Christchurch, New Zealand} 
\author{J. Links} 
\email{jrl@maths.uq.edu.au} 
\affiliation{Department of Mathematics, The University of
Queensland, Brisbane, 4072, Australia}

%\vspace{10pt}
\begin{abstract}
We show that a non-associative structure applied 
to the algebra of Fermi
operators with $su(3)$ colour degrees of freedom 
leads to a consistent Fermi statistic for the tensor
operators of the colour algebra. 
A consequence of this construction is that leads to quark
confinement, without the need to resort to a confining force.
Confinement arises as a symmetry constraint in much the same manner as the
Pauli exclusion principle. 
\end{abstract}

\maketitle  
                     
%************************** Text Begins here ******************************

%  Greek letters
\def\aa{\alpha} 
\def\bb{\beta}
\def\a{\hat a}
\def\b{\hat b}
\def\d{\dagger}
\def\de{\delta} 
\def\e{\epsilon}
\def\ve{\varepsilon}
\def\g{\gamma}
\def\K{\kappa}
\def\ap{\approx}
\def\l{\lambda}
\def\o{\omega}
\def\t{\tilde{\tau}}
\def\s{\sigma}
\def\D{\Delta}
\def\L{\Lambda}
\def\T{{\cal T}}
\def\TT{{\tilde{\cal T}}}
\def\E{{\cal E}} 
\def\f{\overline{f}}
\def\q{\overline{q}}
\def\tp{\otimes}
\def\I{{\rm id}}
\def\rar{\rightarrow}
% Shorthands for \begin{equation} and the like

%\def\beq{\begin{equation}}
%\def\eeq{\end{equation}}
\def\bea{\begin{eqnarray}}
\def\eea{\end{eqnarray}}
\def\ba{\begin{array}}
\def\ea{\end{array}}
\def\no{\nonumber}
\def\le{\langle}
\def\re{\rangle}
\def\lt{\left}
\def\rt{\right}
\def\o{\omega}
\def\d{\dagger}
\def\nn{\nonumber}
\def\j{{ {\cal J}}}
\def\n{{\hat n}}
\def\N{{\hat N}}
\def\A{{\cal A}}
\def\TT{{\tilde {\cal T}}}

\newcommand{\reff}[1]{eq.~(\ref{#1})}

%\newpage
%\vskip.3in
%\begin{multicols}{2}

The current description of the strong nuclear force is provided by a  
non-abelian $su(3)$ gauge theory. This theory is known as
quantum chromodynamics \cite{wilczek}. In this approach the fundamental
particles which experience the strong force are called quarks,  
first proposed by Gell-Mann \cite{gellmann} and Zweig \cite{z}. 
The original introduction of quarks was to give a description of
elementary particles with  
fractional charge which provide a theoretical model for the hadrons.
The hadrons fall into two classes, the baryons which are fermionic and
the mesons which are bosonic. In the quark model, the baryons are
comprised of three quarks and the mesons of a quark and an antiquark. 
Thus each quark is taken to be a fermion, with spin 1/2. The fractional
charges of the quarks are characterised by the flavour degree of 
freedom \cite{gn}.
It is now accepted there exist six quark flavours. It was quickly
realised that within this approach the Pauli exclusion principle was
seen to be violated, since there exist baryons which are described by
three quarks in the same spin and flavour states. A resolution was proposed
\cite{hn,green} to introduce an additional fermionic 
degree of freedom, now known
as colour. The quarks carry colour charge which can take one of three
values; red, green or blue. Transformation of colour are described by
the Lie algebra $su(3)$, and the colour charges provide the mechanism
for the strong interaction in the framework of the $su(3)$ gauge theory. 
One aspect of this theory is that quarks must be confined since to date no
individual quark has been observed \cite{green,dms}. This is thought to be the
consequence of a confining force. In this letter we provide a
basis for quark confinement due solely to colour symmetry. This
arises as a requirement to provide a unique boson or fermion statistic
for the tensor operators of the colour algebra. We will show that this
solution can be achieved by introducing a non-associative structure for
the Fermi algebra which describes the quarks, and confinement is deduced
as a consequence.  

To illustrate our motivation, 
we begin with a Fermi algebra generated by $f_\mu,\,f^\d_\mu$ where
$\mu$ is a colour index taking one of three values. The
usual anti-commutation relations hold  
\bea
&&\{f_\mu,f_\nu\}=\{f^\d_\mu,f^\d_\nu\}=0, \no \\
&&\{f_\mu,f^\d_\nu\}=\delta_{\mu\nu}I \no \eea   
where $I$ is the identity operator.
Setting $E_{\mu\nu}=f^\d_\mu f_\nu$ it is easily shown that 
\bea &&[E_{\mu\nu},\,E_{\rho\omega}]
=\delta_{\nu\rho}E_{\mu\omega}-\delta_{\mu\omega}E_{\rho\nu}\label{cr}
\eea   
thus realising the Lie algebra $u(1)\oplus su(3)$ where the $u(1)$
generator is 
\bea  && C=\sum_{\mu}E_{\mu\mu}. \label{c} \eea     
It can also be shown that 
$$[E_{\mu\nu},\,f^\d_\rho]= \delta_{\nu\rho} f^\d_{\mu}$$ 
so the Fermi creation operators provide an $su(3)$ tensor operator which
transforms as the fundamental representation. Similarly 
$$[E_{\mu\nu},\,f_\rho]= - \delta_{\mu\rho} f_{\nu}$$
indicating that the annihilation operators form a tensor operator
transforming as the dual fundamental representation. Both of these
tensor operators are clearly fermionic. 

Defining 
$$\Gamma_{\mu}=\sum_{\nu,\rho}\ve_{\mu\nu\rho}f^\d_{\nu}f^\d_{\rho},$$ 
where $\ve_{\mu\nu\rho}$ is the Levi-Civita symbol, it can be shown  
that the $\Gamma_{\mu}$ transform as the dual fundamental tensor
while the $\Gamma^\d_{\mu}$ transform as the fundamental tensor.  
Assuming associativity of the Fermi
algebra gives 
\bea &&f^\d_{\mu}f^\d_{\nu}f^\d_{\rho}=-f^\d_{\nu}f^\d_{\mu}f^\d_{\rho}
=f^\d_{\nu}f^\d_{\rho}f^\d_{\mu} \no \eea 
which shows that 
$\Gamma_\mu$  and $\Gamma^\d_{\mu}$ are bosonic in the sense that 
$$ f^\d_\nu\Gamma_\mu=\Gamma_\mu f^\d_\nu, ~~~f_\nu \Gamma_{\mu}^\d=
\Gamma_{\mu}^\d f_\nu. $$ 
%which follows from 
%\bea
%f_\mu^\d(f^\d_\nu f^\d_\rho)&=&(f_\mu^\d f^\d_\nu)f^\d_\rho\no \\
%&=&-(f_\nu^\d f^\d_\mu)f^\d_\rho\no \\
%&=&-f_\nu^\d(f^\d_\mu f^\d_\rho)\no \\
%&=&f_\nu^\d(f^\d_\rho f^\d_\mu)\no \\
%&=&(f_\nu^\d f^\d_\rho)f^\d_\mu.\no
%\eea
Thus, by assuming associativity of the Fermi algebra it is not possible
to define a unique bosonic or fermionic statistic for a tensor operator
which transforms according to a given representation. Specifically, we
require both $\Gamma_{\mu}$ and $\Gamma_{\mu}^\d$ to be fermionic as
they carry a single fermionic colour label. 
We can overcome this contradiction by forsaking
associativity. For example, assuming instead 
\bea && f_\mu^\d(f^\d_\nu f^\d_\rho)=-(f_\mu^\d f^\d_\nu)f^\d_\rho
\label{e0} \eea     
we can repeat the above calculation to show 
\bea
f_\mu^\d(f^\d_\nu f^\d_\rho)&=&-(f_\mu^\d f^\d_\nu)f^\d_\rho\no \\
&=&(f_\nu^\d f^\d_\mu)f^\d_\rho\no \\
&=&-f_\nu^\d(f^\d_\mu f^\d_\rho)\no \\
&=&f_\nu^\d(f^\d_\rho f^\d_\mu)\no \\
&=&-(f_\nu^\d f^\d_\rho)f^\d_\mu\no   
\eea
which then implies $$f^\d_\nu\Gamma_\mu=-\Gamma_\mu f^\d_\nu$$
so the $\Gamma_\mu$ now transform as a fermionic tensor operator. 
Similarly the $\Gamma^\d_{\mu}$ are fermionic and the 
analysis can be extended to show that the colour singlet
$$\Psi=\sum_{\mu}f^\d_{\mu}\Gamma_{\mu} $$
is bosonic. 

Non-associative structures can be rigorously defined using a
category-theoretic approach \cite{mac}. In relation to the $su(3)$
colour algebra this problem was first studied in \cite{joyce-s} from
which quark confinement was deduced. The above results are consistent to
those in \cite{joyce-s}. The necessary mathematical
formalism was developed in \cite{joyce-q,joyce-b,joyce-v}. In \cite{ijl}
an algebraic approach was adopted which allows the theory to be
applied on a general level for any Lie algebra symmetry. The aim of this letter
is to extend the 
above construction for quarks with either colour or anticolour labels 
in such a way that it 
gives a fermionic character for  baryonic states, and 
bosonic for  mesonic states. By introducing antiquarks, the colour
symmetry is enlarged to $u(1)\oplus su(3)$, which we will show
below.  
We will present a construction for this $u(1)\oplus
su(3)$ case following the approach of \cite{ijl}, without too much emphasis
on the technical details, which can be found in
\cite{joyce-s,joyce-q,joyce-b,joyce-v,ijl}. We will see that in this
framework the quarks are confined to hadronic states with the correct
statistics.

Let us begin at the level of an arbitrary non-associative algebra 
$\A$. For $A,\,B,\,C\in \A$ we treat 
the composite operators $A(BC)$ and $(AB)C$ 
as being distinct. We will require however the existence of 
an invertible 
function  $\phi$ which allows us to map between these two choices of
association; i.e. 
$$\phi[(AB)C]=\zeta(A,B,C)[A(BC)]$$ 
where in general $\zeta(A,B,C)$ can take complex values. We also require
a commutativity operator of the form 
$$\s[AB]= \gamma(A,B)BA$$ 
where $\gamma(A,B)$ is complex valued. 
Given a $\s$ which is defined on a generating set for $\A$ 
it can be extended, through use of $\phi$,  
to composite operators in the
manner depicted graphically below 
$$
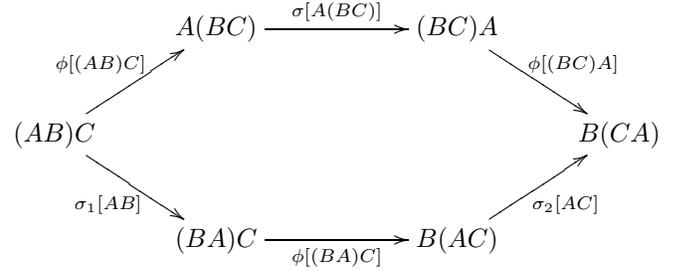
\xymatrix{
								&
A(BC) \ar[rr]^{\s[A(BC)]}                           &&
(BC)A \ar[dr]^{\phi[(BC)A]}                         &
\\
(AB)C \ar[ur]^{\phi[(AB)C]} \ar[dr]_{\s_1[AB]} &&&&
B(CA) \\
&
(BA)C \ar[rr]_{\phi[(BA)C]}                         &&
B(AC) \ar[ur]_{\s_2[AC]}                           &
}
$$
or equivalently 
$$\s[A(BC)]=\phi^{-1}\circ \s_2 \circ \phi \circ 
\s_1 \circ \phi^{-1}[A(BC)], $$  
where $\s_1$ interchanges the first two operators and $\s_2$
interchanges the second and third. 

To apply this approach to a many quark system we let 
$$\{f_{i\mu},f^{\d}_{i\mu},\f_{i\mu},\f^{\d}_{i\mu}:i=1,...,N,\,\mu={\rm
red,~green,~blue}\}$$
be a set of Fermi operators where $f_{i\mu}^\d$ is the creation operator for
a quark with colour $\mu$ and quantum label $i$ which denotes all other
degrees of freedom such as spin, hypercharge etc., and  
$\f^\d_{i\mu}$ is the creation
operator for the corresponding antiquark. We take the usual
anti-commutation relations for these operators
\bea
&&\{f^\d_{i\mu},f^\d_{j\nu}\}=\{\f^\d_{i\mu},\f^\d_{j\nu}\}=0, \no \\
&&\{f_{i\mu},f_{j\nu}\}=\{\f_{i\mu},\f_{j\nu}\}=0, \no \\
&&\{f^\d_{i\mu},\f^\d_{j\nu}\}=\{f_{i\mu},\f_{j\nu}\}=0, \no \\
&&\{f_{i\mu},\f^\d_{j\nu}\}=\{\f_{i\mu},f^\d_{j\nu}\}=0, \no \\
&&\{f_{i\mu},f^\d_{j\nu}\}=\delta_{ij}\delta_{\mu\nu}I \no \\
&&\{\f_{i\mu},\f^\d_{j\nu}\}=\delta_{ij}\delta_{\mu\nu}I \no \eea
and let $F$ denote the full enveloping algebra. We define
$$E_{\mu\nu}=\sum_{i=1}^N(f^\d_{i\mu}f_{i\nu}+\f_{i\mu}\f^\d_{i\nu})$$ 
which satisfy the relations (\ref{cr}). Besides the $u(1)$ invariant
$C$ given by (\ref{c}) there is an additional $u(1)$ symmetry given by 
$$K=\sum_{i=1}^N\sum_\mu (f^\d_{i\mu}f_{i\mu}-\f_{i\mu}\f^\d_{i\mu}). $$ 
Thus the colour symmetry algebra extends to $g=u(1)\oplus su(3)$. 
Let $F^+$ and $F^-$ be the subalgebras of $F$ consisting of all creation
and annihilation operators respectively. Clearly we have 
\bea &&[g,\,F^+]=F^+, ~~~ [g,\,F^-]=F^-  \no \eea 
so the elements of $F^+$ provide a tensor operator for $g$, as do the
elements of $F^-$. 

Next we define the concepts of {\em parity} 
$p(A)\in {\mathbb Z}_2$ and {\em triality} $t(A)\in {\mathbb Z}_3 $ 
of an operator $A$. Let 
$$p[I]=0,~~ p[f_{i\mu}]=p[f^\d_{i\mu}]=p[\f_{i\mu}]=p[\f^\d_{i\mu}]=1$$ 
and 
$$t[I]=0, ~~t[f^\d_{i\mu}]=t[\f_{i\mu}]=1,
~~t[f_{i\mu}]=t[\f^\d_{i\mu}]=2$$ 
for all labels $i$ and $\mu$. For all composite 
operators the parity is determined by 
$$p[AB]=(p[A]+p[B])({\rm mod}\,2) $$
and triality through  
$$t[AB]=(t[A]+t[B])({\rm mod}\,3). $$ 
We define $\phi$ on both $F^+$ and $F^-$ as follows, 
which was derived using the
algebraic method of \cite{ijl}:   
\bea &&\phi[(AB)C] =A(BC) \no \\ 
&&~~~~~~{\rm if}~p[X]=t[X]=0, ~~~X=A,\,B ~{\rm or ~} C, \no \\
&&\phi[(AB)C] =A(BC) \no \\
&&~~~~~~{\rm if}~p[BC]= t[BC]=0,\no \\
%\phi[(AB)C]&=&-A(BC) ~{\rm if}~~p[A]\neq 0,\,t[A]= 0,\,t[B]=t[C]\neq
%0.\no \\
&&\phi[(AB)C]=-A(BC) ~~~~{\rm otherwise.}  
\no  \eea 
The above non-associative coupling falls within the class given in
\cite{joyce-s}. We also define 
$$\sigma[AB]=-BA ~~~{\rm for~ all}~~A,\,
B=f_{i\mu},\,\f_{j\nu}~{\rm ~or~} f^\d_{i\mu},\,\f^\d_{j\nu}. $$ 
Using this non-associative scheme one can check, for example, that 
\bea
\s[f_{i\mu}^\d(f^\d_{j\nu} f^\d_{k\rho})]
&=&\phi^{-1}\circ \s_2 \circ \phi \circ \s_1 \circ \phi^{-1}
[f_{i\mu}^\d(f^\d_{j\nu}f^\d_{k\rho})]\no \\
&=&-\phi^{-1}\circ \s_2 \circ \phi \circ \s_1
[(f_{i\mu}^\d f^\d_{j\nu})f^\d_{k\rho}]\no \\
&=&\phi^{-1}\circ \s_2 \circ \phi [(f_{j\nu}^\d f^\d_{i\mu})f^\d_{k\rho}]
\no \\
&=&-\phi^{-1}\circ \s_2 [f_{j\nu}^\d (f^\d_{i\mu} f^\d_{k\rho})]\no
\\
&=&\phi^{-1}[f_{j\nu}^\d (f^\d_{k\rho} f^\d_{i\mu})]\no
\\
&=&-(f_{j\nu}^\d f^\d_{k\rho}) f^\d_{i\mu} 
\label{e1} \eea
and similarly 
%can be extended to many operator terms such as  
\bea \s[\f^\d_{i\mu}(\f^\d_{j\nu}\f^\d_{k\rho} )]
&=&-(\f^\d_{j\nu}\f^\d_{k\rho} )\f^\d_{i\mu}\label{e2}  
\\
\s[f^\d_{i\mu}(f^\d_{j\nu}\f^\d_{k\rho})]
&=&(f^\d_{j\nu}\f^\d_{k\rho})f^\d_{i\mu}\label{e3} \\
\s[\f^\d_{i\mu}(f^\d_{j\nu}\f^\d_{k\rho})]
&=&(f^\d_{j\nu}\f^\d_{k\rho})\f^\d_{i\mu}. \label{e4}\eea

It will prove useful to define the map 
$$q[(AB)(CD)]=\zeta(A,B,C,D)(AB)(CD) $$ 
in terms of $\phi$ through the diagram 
$$
\xymatrix{
(AB)(CD)   \ar[rr]^{q[(AB)(CD)]}&
&
(AB)(CD) \ar[dd]^{\phi[(AB)(CD)]}
\\
&&\\
((AB)C)D \ar[uu]^{\phi[((AB)C)D]} \ar[dd]_{\phi_1[(AB)C]} &&
A (B (CD))  \\
&&\\
(A(BC))D \ar[rr]_{\phi[(A(BC))D]} &
&
A((BC)D) \ar[uu]_{\phi_2[((BC)D]}
}
$$
where $\phi_1$ acts on the first three operators and $\phi_2$ acts on
the last three operators. 
We can equivalently write 
$$q[(AB)(CD)]=\phi^{-1}\circ \phi_2 \circ \phi \circ \phi_1 \circ 
\phi^{-1}[(AB)(CD)]. $$ 
As an example we consider 
\bea 
&&q[(f^\d_{i\mu}f^\d_{j\nu})( f^\d_{k\rho} f^\d_{l\omega})]
\no \\
&&~~~= 
\phi^{-1}\circ \phi_2 \circ \phi \circ \phi_1 \circ 
\phi^{-1}[(f^\d_{i\mu}f^\d_{j\nu})( f^\d_{k\rho} f^\d_{l\omega})] \no \\
&&~~~=-\phi^{-1}\circ \phi_2 \circ \phi \circ \phi_1  
[((f^\d_{i\mu}f^\d_{j\nu}) f^\d_{k\rho}) f^\d_{l\omega}] \no \\ 
&&~~~=\phi^{-1}\circ \phi_2 \circ \phi   
[(f^\d_{i\mu}(f^\d_{j\nu} f^\d_{k\rho})) f^\d_{l\omega}] \no \\
&&~~~=-\phi^{-1}\circ \phi_2    
[f^\d_{i\mu}((f^\d_{j\nu} f^\d_{k\rho}) f^\d_{l\omega})] \no \\
&&~~~=\phi^{-1}    
[f^\d_{i\mu}(f^\d_{j\nu} (f^\d_{k\rho} f^\d_{l\omega}))] \no \\
&&~~~=-(f^\d_{i\mu}f^\d_{j\nu}) (f^\d_{k\rho} f^\d_{l\omega}).\label{con} \eea 
In fact, we can give a general formula for the action of $q$
which can be derived from \cite{ijl} 
\bea 
&&q[(AB)(CD)]=(AB)(CD) \no \\
&&~~~~~{\rm if} ~p[X]=t[X]=0, ~~~X=A,\,B,\,C~{\rm or}~D, \no \\
&&q[(AB)(CD)]=(AB)(CD) \no \\
&&~~~~~{\rm if}~p[XY]=t[XY]=0 
\no \\
&&~~~~~~~~~{\rm for }~ X=A,\,Y=B ~{\rm or}~ X=C,\,Y=D, \no \\
&&q[(AB)(CD)]=-(AB)(CD) ~~~~{\rm otherwise.}
\label{q} \eea

The next step is to impose that the algebra is invariant under both
$\phi$ and $\s$; i.e.
\bea
\phi[(AB)C]&=&(AB)C ~~~~{\rm for~all} ~A,\,B,\,C \no \\
\s[AB]&=&AB ~~~~{\rm for~all} ~A,\,B. \no \eea
As $q$ is defined solely in terms of $\phi$, invariance under $q$ follows
$$q[(AB)(CD)]=(AB)(CD). $$ 
Doing this allows us to deduce, for instance, from
(\ref{e1},\ref{e2},\ref{e3},\ref{e4})
\bea 
f^\d_{i\mu}(f^\d_{j\nu}f^\d_{k\rho})
&=&-(f^\d_{j\nu}f^\d_{k\rho})f^\d_{i\mu}\no \\
\f^\d_{i\mu}(\f^\d_{j\nu}\f^\d_{k\rho})
&=&-(\f^\d_{j\nu}\f^\d_{k\rho})\f^\d_{i\mu}\no \\
f^\d_{i\mu}(f^\d_{j\nu}\f^\d_{k\rho})
&=&(f^\d_{j\nu}\f^\d_{k\rho})f^\d_{i\mu}\no  \\
\f^\d_{i\mu}(f^\d_{j\nu}\f^\d_{k\rho})
&=&(f^\d_{j\nu}\f^\d_{k\rho})\f^\d_{i\mu} \no \eea
which provides a generalisation of (\ref{e0}) to the present case. 

The constraint of invariance with respect to $\phi$ and $\s$
leads directly to exclusion and confinement. For example
$$ \s[f^\d_{i\mu}f^\d_{i\mu}]=-f^\d_{i\mu}f^\d_{i\mu}, $$ 
therefore 
$$ f^\d_{i\mu}f^\d_{i\mu}=-f^\d_{i\mu}f^\d_{i\mu}, $$  
which can only be true if 
$$(f^\d_{i\mu})^2=0. $$ 
This is precisely the exclusion principle stating that two quarks
cannot occupy the same state. 
Likewise we have from (\ref{con}) 
$$
q[(f^\d_{i\mu}f^\d_{j\nu}) (f^\d_{k\rho} f^\d_{l\omega})]
=-(f^\d_{i\mu}f^\d_{j\nu}) (f^\d_{k\rho} f^\d_{l\omega}), 
$$ 
so that 
$$(f^\d_{i\mu}f^\d_{j\nu}) (f^\d_{k\rho} f^\d_{l\omega})
=-(f^\d_{i\mu}f^\d_{j\nu}) (f^\d_{k\rho} f^\d_{l\omega}), $$ 
implying 
$$(f^\d_{i\mu}f^\d_{j\nu}) (f^\d_{k\rho} f^\d_{l\omega})=0 $$ 
which is true for all choices for the quantum indices $i$ and 
colour index $\mu$. Thus we immediately see that it is not possible to
have a state of four quarks. 

In category theory terms the requirement that $q$ is the identity 
operator means that the category is monoidal \cite{mac}. Thus the
construction that we have provided here gives an example of a category
which is not monoidal. For general discussion and results on such
categories we refer to \cite{joyce-q,joyce-b,joyce-v}. Once we impose
that the algebra is invariant under $\phi$, and consequently under $q$,
we essentially make a projection onto the monoidal subalgebra of the
full algebra. In the present example we can see from (\ref{q}) that the
monoidal subalgebras for both $F^+$ and $F^-$ are the 
subalgebras of operators with triality zero.
It is also clear that these subalgebras {\em are} associative. 
As a generating set for the monoidal subalgebra of $F^+$   
we have the baryonic creation operators    
$$(f^\d_{i\mu}f^\d_{j\nu}) f^\d_{k\rho}, ~~~~
(\f^\d_{i\mu}\f^\d_{j\nu}) \f^\d_{k\rho} $$ 
and the mesonic creation operators   
$$f^\d_{i\mu}\f^\d_{j\nu}. $$ 
%the quark-quark and antiquark-antiquark exchange operators   
%$$f^\d_{i\mu}f_{j\nu},~~~~ \f^\d_{i\mu}\f_{j\nu} $$  
%and the quark-antiquark exchange operators
%$$(f^\d_{i\mu}f^\d_{j\nu}) \f_{k\rho}, ~~~~ (\f^\d_{i\mu}\f^\d_{j\nu})
%f_{k\rho} $$  
%$$(f^\d_{i\mu}\f_{j\nu}) \f_{k\rho}, ~~~~ (\f^\d_{i\mu}f_{j\nu})
%f_{k\rho}. $$  
The generating set for the monoidal subalgebra of $F^-$ is given by the
hermitian conjugates of the above operators. 
All other operators of triality zero for either $F^+$ and $F^-$  
can be expressed as products of elements of these generating sets. 
Thus we conclude the quarks are confined to hadronic states, that is a
baryonic state of three quarks or three antiquarks, 
or a mesonic state of a quark and an
antiquark. It is worth mentioning finally that one can check that  
\bea 
&&((f^\d_{i\mu}f^\d_{j\nu}) f^\d_{k\rho})
((f^\d_{l\omega}f^\d_{m\alpha}) f^\d_{n\beta})
\no \\ 
&&~~~~=-((f^\d_{l\omega}f^\d_{m\alpha}) f^\d_{n\beta})
((f^\d_{i\mu}f^\d_{j\nu}) f^\d_{k\rho}) \no \\ 
&&((\f^\d_{i\mu}\f^\d_{j\nu}) \f^\d_{k\rho})
((\f^\d_{l\omega}\f^\d_{m\alpha}) \f^\d_{n\beta})
\no \\
&&~~~~=-((\f^\d_{l\omega}\f^\d_{m\alpha}) \f^\d_{n\beta})
((\f^\d_{i\mu}\f^\d_{j\nu}) \f^\d_{k\rho}) \no \\
&&((f^\d_{i\mu}f^\d_{j\nu}) f^\d_{k\rho})
((\f^\d_{l\omega}\f^\d_{m\alpha}) \f^\d_{n\beta})
\no \\
&&~~~~=-((\f^\d_{l\omega}\f^\d_{m\alpha}) \f^\d_{n\beta})
((f^\d_{i\mu}f^\d_{j\nu}) f^\d_{k\rho}) \no \\
&&((f^\d_{i\mu}f^\d_{j\nu}) f^\d_{k\rho})
(f^\d_{l\omega}\f^\d_{m\alpha})
=(f^\d_{l\omega}\f^\d_{m\alpha})
((f^\d_{i\mu}f^\d_{j\nu}) f^\d_{k\rho}) \no \\
&&((\f^\d_{i\mu}\f^\d_{j\nu}) \f^\d_{k\rho})
(f^\d_{l\omega}\f^\d_{m\alpha})
=(f^\d_{l\omega}\f^\d_{m\alpha})
((\f^\d_{i\mu}\f^\d_{j\nu}) \f^\d_{k\rho}) \no \\
&&(f^\d_{i\mu}\f^\d_{j\nu})(f^\d_{k\rho} \f^\d_{l\omega})
=(f^\d_{k\rho} \f^\d_{l\omega})(f^\d_{i\mu}\f^\d_{j\nu}) \no \eea 
so the construction also gives the correct statistics for the hadrons,
i.e. baryons are fermionic and mesons are bosonic. 

%\vskip.3in
%\acknowledgments

PSI is a Postdoctoral Fellow supported by the Japanese Society for the
Promotion of Science.
WPJ acknowledges the support of the New Zealand Foundation for Research,
Science and Technology (contract number UOCX0102).
JL thanks the Graduate School of Mathematical
Sciences, The University of Tokyo and the Department of Physics and
Astronomy, The University of Canterbury for generous hospitality and
acknowledges financial support from the Australian Research
Council.
%\newpage
%\vskip.3in
 
%\end{multicols}
\end{document}